\RequirePackage{ifpdf}
\ifpdf % We are running pdfTeX in pdf mode
\documentclass[pdftex]{sigma}
\else
\documentclass{sigma}
\fi

\begin{document}

\allowdisplaybreaks

\renewcommand{\PaperNumber}{088}

\FirstPageHeading

\ShortArticleName{Flatland Position-Dependent-Mass}

\ArticleName{Flatland Position-Dependent-Mass:\\ Polar Coordinates,
Separability and Exact Solvability}

\Author{S. Habib MAZHARIMOUSAVI and Omar MUSTAFA}

\AuthorNameForHeading{S.H.~Mazharimousavi and O.~Mustafa}

\Address{Department of Physics, Eastern Mediterranean University,\\
 G Magusa, North Cyprus, Mersin 10, Turkey}
\Email{\href{mailto:habib.mazhari@emu.edu.tr}{habib.mazhari@emu.edu.tr}, \href{mailto:omar.mustafa@emu.edu.tr}{omar.mustafa@emu.edu.tr}}

\ArticleDates{Received August 15, 2010, in f\/inal form October 26, 2010;  Published online October 29, 2010}

\Abstract{The kinetic energy operator with position-dependent-mass in plane polar coordinates is obtained. The
 separability of the corresponding Schr\"odinger equation is discussed. A hypothetical toy model is reported and two
exactly solvable examples are studied.}


\Classification{81Q05; 81Q60}

\section{Introduction}

Position-dependent-mass (PDM) quantum particles have attracted research
attention over the last few decades \cite{1,2,3,4,5,6,7,8,9,10,11,12,13,14,15,16,17,
18,19,20,21,22,23,24,25,26,27,28,29}. Such attention is inspired not
only by the PDM feasible applicability in the study of various physical
problems (e.g., many-body problem, semiconductors, quantum dots, quantum
liquids, etc.), but also by the mathematical challenge associated with the
corresponding von Roos Hamiltonian. The non-commutativity between the
momentum operator and the position-dependent-mass introduces an ordering
ambiguity in the kinetic energy operator
\begin{gather}\label{eq1}
T=-\frac{\hbar ^{2}}{4}\big\{ m( \vec{r}) ^{\gamma }\vec{\nabla}
m( \vec{r}) ^{\beta }\mathbf{\cdot }\vec{\nabla}m( \vec{r}) ^{\alpha }+m( \vec{r}) ^{\alpha } \vec{\nabla}m(
\vec{r}) ^{\beta }\mathbf{\cdot }\vec{\nabla}m( \vec{r})
^{\gamma}\!\big\} .
\end{gather}
Obviously, changing the values of $\alpha$, $\beta $, and $\gamma $ would
change $T$. Hence, $\alpha $, $\beta $, and $\gamma $ are called the
ordering ambiguity parameters subjected to von Roos constraint $\alpha
+\beta +\gamma =-1$ (cf., e.g., \cite{25,26,27,28,29}). In the literature, one may f\/ind
the orderings of Gora and William ($\beta =\gamma =0$, $\alpha =-1$), Ben
Daniel and Duke ($\alpha =\gamma =0$, $\beta =-1$), Zhu and Kroemer ($\alpha
=\gamma =-1/2$, $\beta =0$), Li and Kuhn ($\beta =\gamma =-1/2$, $\alpha =0$), and Mustafa and Mazharimousavi
($\alpha =\gamma =-1/4$, $\beta =-1/2$)
(cf., e.g., \cite{10, 29} for more details on this issue). However, it has been
observed (cf., e.g.,~\cite{29}) that the physical and/or mathematical
admissibility of a given ambiguity parameters set depends not only on the
continuity conditions at the abrupt heterojunction boundaries but also on
the position-dependent-mass and/or potential forms. The general consensus is
that there is no unique choice for these ambiguity parameters.

On the other hand, the fabrication of the essentially quasi-zero-dimensional
quantum systems (like quantum dots (QD)) that are populated by two
dimensional f\/latland quantum particles (electrons in the QDs case) conf\/ined
by an artif\/icial potential has inspired intensive research activities. Under
such dimensional settings, a quantum particle endowed with a
position-dependent-mass, $m( \vec{r}) =m_{\circ }M( \vec{r}) =m_{\circ }M( \rho ,\varphi ) $,  would be an interesting
case to study, therefore. To the best of our knowledge, such PDM settings
have never been discussed in the literature, within our forthcoming
methodical proposal at least.

For the sake of separability, we recollect (in Section~\ref{section2}) the essentials of
the kinetic ener\-gy operator in plane polar coordinates mandated by a
position dependent mass of the form $M( \rho ,\varphi ) =g(
\rho )  f( \varphi ) $, $g( \rho) =\rho ^{-2}$
and an interaction potential $V( \rho ,\varphi ) =\tilde{V}(
\rho ) /f( \varphi ) $. In Section~\ref{section3}, we show that whilst
the radial part (in~\eqref{eq12} below) can be brought into the one-dimensional Schr\"{o}dinger format through a simple choice $R( \rho ) =\rho
^{-3/2}\ U( \rho ) $, a point canonical transformation (PCT) is
needed for the azimuthal angular part (see~\eqref{eq13} below). Moreover, a~hypothetical toy model with $M( \rho ,\varphi ) =\rho ^{-2}\cos
^{2}\varphi $, $\rho \in ( 0,\infty ) $, $\varphi \in (
0,2\pi ) ,$ and $V( \rho ,\varphi ) =-V_{\circ }\rho
^{2k}/2\cos ^{2}\varphi $, $V_{\circ }>0$, is considered (in the same
section). Two constructive exactly-solvable toy examples of fundamental
nature, with $m( \rho ,\varphi ) =1/\rho ^{2}$, are given in
Section~\ref{section4}. Namely and ef\/fectively, a Coulomb-like and a
harmonic-oscillator-like radial Schr\"{o}dinger models. We conclude in
Section~\ref{section5}.

\section{Essentials of the kinetic energy operator\\ in plane polar
coordinates and separability}\label{section2}

The kinetic energy operator $T$, in \eqref{eq1}, for a PDM quantum particle with $%
m( \vec{r}) =M( \rho ,\varphi ) $, moving in a
two-dimensional f\/latland (with $\hbar =m_{\circ }=1$ units), reads%
\begin{gather*}%\label{eq2}
T=-\frac{1}{4}\big\{ M( \rho ,\varphi ) ^{\gamma }\vec{\nabla}%
M( \rho ,\varphi ) ^{\beta }\mathbf{\cdot }\vec{\nabla}M(
\rho ,\varphi ) ^{\alpha }+M( \rho ,\varphi ) ^{\alpha }%
\vec{\nabla}M( \rho ,\varphi ) ^{\beta }\mathbf{\cdot }\vec{\nabla%
}M( \rho ,\varphi ) ^{\gamma }\!\big\} .
\end{gather*}%
This would, with $M( \rho ,\varphi ) \equiv M$ (for economy of
notation) and
\begin{gather*}
\vec{\nabla}=\mathbf{\hat{\rho}}\ \partial _{\rho }+\mathbf{\hat{\varphi}}%
\frac{1}{\rho }\partial _{\varphi } \quad \Longrightarrow \quad \vec{\nabla}\cdot \vec{%
\nabla}=\nabla ^{2}=\frac{1}{\rho }\ \partial _{\rho }( \rho \ \partial
_{\rho }) +\frac{1}{\rho ^{2}}\partial _{\varphi }^{2},
\end{gather*}
yield (in a straightforward though rather tedious manner)%
\begin{gather}\label{eq3}
T=-\frac{1}{2}\left[ 2W( \rho ,\varphi ) +\left( \frac{1}{\rho \ M%
}-\frac{M_{\rho }}{M^{2}}\right) \partial _{\rho }-\frac{M_{\varphi }}{\rho
^{2}M^{2}}\partial _{\varphi }+\frac{1}{M}\partial _{\rho }^{2}+\frac{1}{%
\rho ^{2}M}\partial _{\varphi }^{2}\right] ,
\end{gather}%
where%
\begin{gather*}%\label{eq4}
W( \rho ,\varphi ) =\frac{1}{4}\left[ \frac{\xi }{M^{3}}\left(
M_{\rho }^{2}+\frac{M_{\varphi }^{2}}{\rho ^{2}}\right) +\frac{\left( \alpha
+\gamma \right) }{M^{2}}\left( \frac{M_{\rho }}{\rho }+M_{\rho \rho }+\frac{%
M_{\varphi \varphi }}{\rho ^{2}}\right) \right] ,
\\
%\label{eq5}
\xi =\alpha ( \alpha -1) +\gamma ( \gamma -1) -\beta
( \beta +1) ,
\end{gather*}
and
\begin{gather*}
M_{\rho }=\partial _{\rho }M( \rho ,\varphi ) , \qquad M_{\rho \rho
}=\partial _{\rho }^{2}M( \rho ,\varphi ) , \qquad M_{\varphi }=\partial
_{\varphi }M( \rho ,\varphi ) , \qquad M_{\varphi \varphi }=\partial
_{\varphi }^{2}M( \rho ,\varphi ) .
\end{gather*}
A recollection of the time-independent Schr\"{o}dinger equation, for the PDM
quantum particle with $m( \vec{r}) =M( \rho ,\varphi )
$ moving under the inf\/luence of a two-dimensional f\/latland potential $%
V( \rho ,\varphi ) $, implies that%
\begin{gather*}%\label{eq6}
T\psi ( \rho ,\varphi ) +V( \rho ,\varphi ) \psi
( \rho ,\varphi ) =E\psi ( \rho ,\varphi ) ,
\end{gather*}%
where $T$ is given in \eqref{eq3}. The separation of variables of which suggests a
wave function of the form
\begin{gather*}
\psi ( \rho ,\varphi ) =R( \rho ) \Phi ( \varphi
),\qquad  \rho \in ( 0,\infty ) ,\qquad \varphi \in ( 0,2\pi
) ,
\end{gather*}
to obtain%
\begin{gather}
\left( \frac{1}{\rho M}-\frac{M_{\rho }}{M^{2}}\right)\! \frac{\partial
_{\rho }R}{R} - \frac{M_{\varphi }}{\rho ^{2}M^{2}}\frac{\partial _{\varphi
}\Phi }{\Phi } + \frac{1}{M}\frac{\partial _{\rho }^{2}R}{R} + \frac{1}{\rho
^{2}M}\frac{\partial _{\varphi }^{2}\Phi }{\Phi }=-2[ E - V( \rho
,\varphi )  + W( \rho ,\varphi ) ] .\!\!\label{eq7}
\end{gather}%
In the search of feasible separability of this equation, it is obvious that
a mass setting of the form%
\begin{gather}\label{eq8}
M( \rho ,\varphi ) =g( \rho)   f( \varphi )
\end{gather}
would, with%
\begin{gather*}
g( \rho ) =\rho ^{-2},
\end{gather*}%
ease separability and allow $W( \rho ,\varphi ) \longrightarrow
\tilde{W}( \varphi ) $, where%
\begin{gather}\label{eq9}
\tilde{W}( \varphi ) =\frac{1}{4}\left[ \xi \left( \frac{4}{f( \varphi ) }+\frac{[ \partial _{\varphi }f( \varphi
) ] ^{2}}{f( \varphi ) ^{3}}\right) +\frac{(
\alpha +\gamma ) }{f( \varphi ) }\left( 4+\frac{\partial
_{\varphi }^{2}f( \varphi ) }{f( \varphi ) }\right) \right] .
\end{gather}
Moreover, if the position-dependent-mass $M( \rho ,\varphi ) $
in \eqref{eq8} is interrelated with the interaction potential $V( \rho ,\varphi
) $ through $f( \varphi ) $ of \eqref{eq8} so that%
\begin{gather*}%\label{eq10}
V( \rho ,\varphi ) =\frac{\tilde{V}( \rho ) }{f(\varphi ) },
\end{gather*}%
one may recast equation \eqref{eq7} as
\begin{gather}\label{eq11}
3\rho \frac{\partial _{\rho }R}{R}+\rho ^{2}\frac{\partial _{\rho }^{2}R}{R}%
-2\tilde{V}( \rho ) =\frac{\partial _{\varphi }f( \varphi
) }{f( \varphi ) }\frac{\partial _{\varphi }\Phi }{\Phi }-%
\frac{\partial _{\varphi }^{2}\Phi }{\Phi }-2f( \varphi ) \big[
E+\tilde{W} ( \varphi ) \big] =\lambda ,
\end{gather}%
where $\lambda $ is the separation constant. The separability of \eqref{eq11} is
obvious, therefore. That is, the radial equation reads
\begin{gather}\label{eq12}
\big[ \rho ^{2}\partial _{\rho }^{2}+3\rho   \partial _{\rho }-2\tilde{V}
( \rho ) \big] R ( \rho ) =\lambda R ( \rho
 ) ,
\end{gather}%
and the azimuthal angular equation reads%
\begin{gather}\label{eq13}
\left\{ -\partial _{\varphi }^{2}+\frac{\partial _{\varphi }f ( \varphi
 ) }{f ( \varphi  ) }\partial _{\varphi }-2f ( \varphi
 ) \big[ E+\tilde{W} ( \varphi  ) \big] \right\} \Phi
 ( \varphi  ) =\lambda \Phi  ( \varphi  ) .
\end{gather}

\section{Corresponding one-dimensional Schr\"{o}dinger equations}\label{section3}

It is obvious that, the substitution $R ( \rho  ) =\rho ^{j}
U ( \rho  )$, $j=-3/2$, would eliminate the f\/irst-order derivative
in equation \eqref{eq12} to obtain
\begin{gather}\label{eq14}
\left\{ -\partial _{\rho }^{2}+\left[ \frac{ ( 3/4+\lambda  ) }{%
\rho ^{2}}+\frac{2\tilde{V} ( \rho  ) }{\rho ^{2}}\right] \right\}
U ( \rho  ) =0,\qquad \rho \in  ( 0,\infty  ) .
\end{gather}%
Which suggests that the ef\/fective radial potential is
\begin{gather*}%\label{eq15}
\tilde{V}_{\rm ef\/f} ( \rho  ) =\frac{ ( 3/4+\lambda  ) }{\rho
^{2}}+\frac{2\tilde{V} ( \rho  ) }{\rho ^{2}}.
\end{gather*}%
On the other hand, a point canonical transformation%
\begin{gather*}%\label{eq16}
q' ( \varphi  ) =\partial _{\varphi }q ( \varphi
 ) =\sqrt{f ( \varphi ) }.
\end{gather*}%
with the substitution%
\begin{gather*}%\label{eq17}
\Phi  ( \varphi ) =f ( \varphi  ) ^{1/4}\chi  (
q ( \varphi  )  ) ,
\end{gather*}%
would transform the azimuthal angular equation \eqref{eq13} into%
\begin{gather}\label{eq18}
-\frac{1}{2}\chi ^{\prime \prime } ( q ) +W_{\rm ef\/f} ( q )
\chi  ( q ) =E\chi  ( q ),
\end{gather}%
where
\begin{gather}
W_{\rm ef\/f} ( q )   =-\tilde{W} ( \varphi  ) +\frac{7 [
\partial _{\varphi }f ( \varphi  )  ] ^{2}}{32  f (
\varphi  ) ^{3}}-\frac{\partial _{\varphi }^{2}f ( \varphi  )
}{8f ( \varphi  ) ^{2}}-\frac{\lambda }{2f ( \varphi  ) }
\notag \\
\phantom{W_{\rm ef\/f} ( q )}{}
 =\frac{ [ \partial _{\varphi }f ( \varphi  )  ] ^{2}}{%
32  f ( \varphi  ) ^{3}} ( 7-8\xi  ) -\frac{\partial
_{\varphi }^{2}f ( \varphi ) }{8f ( \varphi ) ^{2}}%
 ( 1+2 ( \alpha +\gamma  )  )
  -\frac{1}{f ( \varphi ) }\left( \xi +\alpha +\gamma +\frac{%
\lambda }{2}\right) .\label{eq19}
\end{gather}
Few illustrative toy examples are in order.

\subsection{A hypothetical toy model}\label{section3.1}

In this toy model we choose to work on a quantum particle endowed with a
position dependent mass
\begin{gather*}%\label{eq20}
M ( \rho ,\varphi ) =\rho ^{-2}\cos ^{2}\varphi,\qquad
\rho \in  ( 0,\infty ) ,\qquad \varphi \in  ( 0,2\pi ) ,
\end{gather*}%
under the inf\/luence of%
\begin{gather*}%\label{eq21}
V ( \rho ,\varphi  ) =-V_{\circ }\frac{\rho ^{2k}}{2\cos
^{2}\varphi },\qquad V_{\circ }>0.
\end{gather*}%
This would ef\/fectively mean that $g ( \rho  ) =\rho ^{-2}$, $
f ( \varphi ) =\cos ^{2}\varphi $, and $\tilde{V} ( \rho
 ) =-V_{\circ }\rho ^{2k}/2$. Although such  a toy model may not relate
to any practical importance, it could be theoretically and/or mathematically
appealing.

To deal with the radial part in equation \eqref{eq14}, we may recollect that the
Bessel's equation $\rho ^{2}Z_{n}^{\prime \prime } ( \rho  ) +\rho
Z_{n}^{\prime } ( \rho ) + ( \rho ^{2}-n^{2} )
Z_{n} ( \rho ) =0$ with $Z_{n} ( \rho ) =U_{n} (
\rho  ) /\sqrt{\rho }$ collapses into
\begin{gather*}%\label{eq22}
\left\{ \partial _{\rho }^{2}+\left[ 1+\frac{\left( 1-4n^{2}\right) }{4\rho
^{2}}\right] \right\} U ( \rho  ) =0.
\end{gather*}%
Which, when compared with \eqref{eq14}, suggests that $V_{\circ }=1$, $k=1$, and $%
\lambda =n^{2}-1$ are feasibly admissible parametric settings for a
physically acceptable well-behaved solution in the form of Bessel functions,
i.e.,
\begin{gather*}%\label{eq23}
U_{n} ( \rho  ) =A\sqrt{\rho }\ J_{n} ( \rho  )
\quad \Rightarrow \quad R_{n} ( \rho ) =\frac{A}{\rho }\ J_{n} ( \rho
 ) .
\end{gather*}

On the other hand, the ef\/fective potential \eqref{eq19} of the azimuthal angular
part of \eqref{eq18}, with $q^{\prime } ( \varphi  ) =\sqrt{f ( \varphi
 ) }\Longrightarrow q ( \varphi  ) =\sin \varphi $, reads
\begin{gather}\label{eq24}
W_{\rm ef\/f} ( q ) =\frac{\zeta _{1}\sin ^{2}\varphi -\zeta _{2}}{\cos
^{4}\varphi }=\frac{\zeta _{1}q^{2}-\zeta _{2}}{\left( 1-q^{2}\right) ^{2}},
\end{gather}%
where%
\begin{gather}\label{eq25}
\zeta _{1}=\frac{3}{8}+\frac{\lambda }{2},
\end{gather}%
and%
\begin{gather*}%\label{eq26}
\zeta _{2}=\frac{\lambda }{2}-\frac{1}{4}+\alpha \left( \alpha -\frac{1}{2}%
\right) +\gamma \left( \gamma -\frac{1}{2}\right) -\beta  ( \beta
+1 ) .
\end{gather*}%
At this point, one may wish to set $W_{\rm ef\/f} ( q ) =0$ by requiring $%
\zeta _{1}=0=\zeta _{2}$. This would yield that $\lambda =-3/4$ (hence $n=1/2
$). Introducing (in addition to von Roos constraint $\alpha +\beta +\gamma
=-1$) yet another constraint on the ordering ambiguity parameters,
therefore. That is,
\begin{gather}\label{eq27}
\frac{5}{8}=\alpha \left( \alpha -\frac{1}{2}\right) +\gamma \left( \gamma -%
\frac{1}{2}\right) -\beta  ( \beta +1 ) .
\end{gather}%
In this case, the azimuthal angular part of the wave function would, with $%
\chi ( q ( \varphi  )  ) =\exp ( imq ) =\exp
 ( im\sin \varphi ) ,$ read
\begin{gather*}
\Phi  ( \varphi  ) =f ( \varphi  ) ^{1/4}\chi  (
q ( \varphi ) ) =\sqrt{\cos \varphi }e^{im\sin \varphi },
\end{gather*}%
where the energy eigenvalues are given as $E=E_{m}=m^{2}/2$ ($m=0,\pm 1,\pm
2,\dots $ is the magnetic quantum number). Hereby, we notice that the only
available quantum state is the one with an irrational radial quantum number $%
n=1/2$ (which can be thought of as $n=n_{\rho }+1/2$, where $n_{\rho }$ is
the regular radial quantum number, $n_{\rho }=0$ in this particular case).
Hence, the radial wave function is $R ( \rho  ) =C J_{1/2} (
\rho  ) /\rho $. Moreover, it should be noted that the constraint in
\eqref{eq27} is satisf\/ied when $\alpha =\gamma =-1/4$, $\beta =-1/2$ (i.e.,
Mustafa's and Mazharimousavi's ordering~\cite{10}). This should not necessarily
mean that the other orderings available in the literature are no good (cf.,
e.g.,~\cite{10}).

However, should we choose $\zeta _{1}\neq 0\neq \zeta _{2}$ in $%
W_{\rm ef\/f} ( q ) $ of \eqref{eq24}, equation \eqref{eq18} would then admit a~solution
in terms of Heun Conf\/luent functions, ${\rm HeunC}$, as%
\begin{gather*}
\Phi  ( \varphi  )   =f ( \varphi  ) ^{1/4}\chi  (
q ( \varphi ) )
  =\sqrt{\cos \varphi }\left[  ( i\cos \varphi  ) ^{1+\sqrt{2\zeta
_{1}-2\zeta _{2}+1}}\right]  \notag \\
\phantom{\Phi  ( \varphi )   =}{}
\times \left\{ C_{1}{\rm HeunC}\left( 0,-\frac{1}{2},\sqrt{2\zeta _{1}-2\zeta
_{2}+1},\frac{E}{2},-\frac{E}{2}-\frac{\zeta _{2}}{2}+\frac{1}{2},\sin
^{2}\varphi \right) \right.  \notag \\
\left.\phantom{\Phi \left( \varphi \right)   =}{}
  + C_{2}\sin \varphi \,{\rm HeunC}\left( 0,\frac{1}{2},\sqrt{2\zeta
_{1}-2\zeta _{2}+1},\frac{E}{2},-\frac{E}{2}-\frac{\zeta _{2}}{2}+\frac{1}{2}%
,\sin ^{2}\varphi \right) \right\} ,%\label{eq28}
\end{gather*}%
Which is to be subjected to the boundary conditions $\Phi  ( \varphi
 ) =\Phi  ( \varphi +2\pi ) $, indeed. Here, $C_{1}$~and~$%
C_{2}$ are constants to be determined. Under such settings, moreover, it is
obvious that one should choose a value of~$E$ (consequently~$\zeta _{2}$ and
$\zeta _{1}$ would be determined for specif\/ic values of the ordering
ambiguity parameters) to f\/ind some value for $\lambda $. Nevertheless,
labeling/classifying the quantum states (i.e., $1s, 2s, 2p, \dots $, etc.)
seems to be non-negotiable and lost in the jam of the multi-parametric
settings, manifested by not only the chosen position dependent mass and
potential but also by the ordering ambiguity parameters as well. However,
such experimental toy model inspires the exploration of the forthcoming
interesting toy examples.

\section[Constructive, exactly-solvable two toy examples with $m(\rho ,\varphi ) =1/\rho ^{2}$]{Constructive, exactly-solvable two toy examples\\ with $\boldsymbol{m(\rho ,\varphi ) =1/\rho ^{2}}$}
\label{section4}

In this section, we consider two examples of exactly-solvable nature (many
other models do amply exist). Strictly speaking, a quantum particle endowed
with a position dependent mass with $g ( \rho ) =1/\rho ^{2}$ and $f ( \varphi  ) =1$ (hence $M ( \rho ,\varphi  ) =1/\rho
^{2}$) would, in ef\/fect, collapse equation \eqref{eq9} into
\begin{gather*}%\label{eq29}
\tilde{W} ( \varphi ) =\alpha ^{2}+\gamma ^{2}-\beta ( \beta
+1 ) ,
\end{gather*}%
and simplify equation \eqref{eq13} to read
\begin{gather*}%\label{eq30}
\big\{ {-}\partial _{\varphi }^{2}-\big[ 2E+2\tilde{W} ( \varphi )
+\lambda \big] \big\} \Phi  ( \varphi ) =0,
\end{gather*}%
where $\Phi  ( \varphi  ) $ should be a single-valued function
(i.e., $\Phi  ( \varphi  ) =\Phi  ( \varphi +2\pi  ) $).
Therefore, it takes the form $\Phi  ( \varphi  ) =e^{im\varphi }$.
In this case, $2E+2\tilde{W} ( \varphi  ) +\lambda =m^{2}$ and
\begin{gather}\label{eq31}
E=\frac{m^{2}-\lambda }{2}-\left[ \alpha ^{2}+\gamma ^{2}-\beta  ( \beta
+1 ) \right] .
\end{gather}

Moreover, we now consider that such a PDM quantum particle is moving in a
force f\/ield described through%
\begin{gather*}%\label{eq32}
\tilde{V} ( \rho  ) =\frac{1}{2}\omega ^{2}\rho ^{2}-\rho .
\end{gather*}%
One would then recast equation \eqref{eq14} as%
\begin{gather}\label{eq33}
\left\{ -\partial _{\rho }^{2}+\left[ \frac{\ell ^{2}-1/4}{\rho ^{2}}+\omega
^{2}-\frac{2}{\rho }\right] \right\} U ( \rho ) =0,
\end{gather}%
where $\ell = \vert m \vert $ and $\ell ^{2}-1/4=3/4+\lambda
\Longrightarrow \ell = ( \lambda +1 ) ^{1/2}$. Hereby, it should be
noted that equation \eqref{eq33} is a two-dimensional Coulombic-like Schr\"{o}dinger
equation. As such, its solution can be inferred from the well-known
two-dimensional Coulombic one. The eigenvalues of which read%
\begin{gather*}%\label{eq34}
\omega ^{2}= ( n_{\rho }+\ell +1 ) ^{-2}\quad \Longrightarrow \quad \omega
= ( n_{\rho }+\ell +1 ) ^{-1}=\big( n_{\rho }+\sqrt{\lambda +1}%
+1\big) ^{-1},
\end{gather*}%
and yield (with $\omega =1/b$ for simplicity)%
\begin{gather*}%\label{eq35}
\lambda = ( b-n_{\rho }-1 ) ^{2}-1,
\end{gather*}%
where $n_{\rho }=0,1,2,\dots $ is the radial quantum number. In this case,
equation \eqref{eq31} reads the energy eigenvalues as%
\begin{gather}\label{eq36}
E_{n_{\rho },m}=\frac{1}{2}\big[ m^{2}- ( b-n_{\rho }-1 ) ^{2}+1%
\big] -\big[ \alpha ^{2}+\gamma ^{2}-\beta  ( \beta +1 ) \big]
.
\end{gather}

Next, we consider the same PDM quantum particle moving now in a radial
potential (repla\-cing that in \eqref{eq33}, in ef\/fect) described by
\begin{gather*}%\label{eq37}
\tilde{V} ( \rho  ) =\frac{a^{2}}{8}\rho ^{4}-\frac{d}{2}\rho ^{2}.
\end{gather*}%
Under such setting, equation \eqref{eq14} reads%
\begin{gather*}%\label{eq38}
\left\{ -\partial _{\rho }^{2}+\left[ \frac{\ell ^{2}-1/4}{\rho ^{2}}+\frac{%
a^{2}}{4}\rho ^{2}-d\right] \right\} U ( \rho  ) =0,
\end{gather*}%
which is ef\/fectively a two-dimensional harmonic-oscillator-like problem that
admits eigenvalues of the form%
\begin{gather*}%\label{eq39}
d=a\big( 2n_{\rho }+\sqrt{\lambda +1}+1\big)\quad \Longrightarrow \quad\lambda
=\left( \frac{d}{a}-2n_{\rho }-1\right) ^{2}-1.
\end{gather*}%
Therefore, the energy eigenvalues for such an oscillator-like problem are%
\begin{gather}\label{eq40}
E_{n_{\rho },m}=\frac{1}{2}\left[ m^{2}-\left( \frac{d}{a}-2n_{\rho
}-1\right) ^{2}+1\right] -\left[ \alpha ^{2}+\gamma ^{2}-\beta  ( \beta
+1 ) \right] .
\end{gather}%
Obviously, the degeneracies associated with the magnetic quantum number ($m=0,\pm 1,\pm 2,\dots $) in the energies \eqref{eq36} and \eqref{eq40} are unavoidably in
point. To remove such degeneracies, a magnetic f\/ield applied perpendicular
to the current two dimensional f\/latland settings (through the symmetric
gauge $\mathbf{A}=\mathbf{B} ( -y,x,0 ) /2$ that may introduce a
term $\sim mB$ to be added to $\lambda $ (cf., e.g., Mustafa~\cite{30}) could be
sought as a remedy. However, such a study readily lies far beyond our
current methodical proposal. Yet, additional degeneracies associated with
the ordering ambi\-guity parameters~$\alpha $ and~$\gamma $ (but not~$\beta $)
are observed. Moreover, within similar parametric changes, one can also get
the corresponding wave functions.

\section{Concluding remarks}\label{section5}

The kinetic energy operator for a quantum particle endowed with a position
dependent mass is a problem with many aspects that are yet to be explored.
In the current work, we tried to study this problem within the context of
the f\/latland plane-polar coordinates $( \rho ,\varphi ) $. In due
course, the essentials of the kinetic energy operator in plane-polar
coordinates are reported. The separability of the related Schr\"{o}dinger
equation is sought through a position dependent mass $M( \rho ,\varphi
) =g( \rho )   f( \varphi ) $, accompanied by an
interaction potential $V( \rho ,\varphi ) =\tilde{V}( \rho
) /f( \varphi) $. Such a~combination is not a unique one.

In the light of our experience above, some observations are in order.

Apart from its practical applicability and multi-parametric complexity
sides, the hypothetical toy model ($M( \rho ,\varphi ) =\rho
^{-2}\cos ^{2}\varphi $ and $V( \rho ,\varphi) =-V_{\circ }\rho
^{2k}/2\cos ^{2}\varphi $) has provided a road-map on the possible technical
dif\/f\/iculties. We have observed that for a physically acceptable radial
solution, for \eqref{eq14}, one may consider $V_{\circ }=1$, $k=1$, and $\lambda
=n^{2}-1$ as proper parametric settings. Moreover, for the azimuthal angular
part with $\zeta _{1}=0=\zeta _{2}$ in \eqref{eq24}, we have found that there is
only one quantum state with a radial quantum number $n_{\rho }=0$ ($n=1/2$)
and a~corresponding magnetic quantum number $m=0,\pm 1,\pm 2,\dots $.
However, for $\zeta _{1}\neq 0\neq \zeta _{2}$, the problem becomes more
complicated and one may solve it following the simplest possible way. For
example, instead of choosing $\lambda $ of \eqref{eq25} and calculate~$E$ of~\eqref{eq18},
one may choose a value for~$E$ f\/irst and calculate $\lambda $ (a~process
that is contemplated to indulge some graphical estimations). Indeed, the
quantum states are readily there but we were unable to properly
identify/classify them (within the context of the well known quantum
numbers). Such a ``quasi-quantum-miss'', say, did not repeat itself for the
rest of the examples reported above (i.e., the quantum states were very well
identif\/ied/classif\/ied within the known quantum numbers as documented in
Section~\ref{section4}).

Moreover, we have shown that for a quantum particle with $m( \rho
,\varphi ) =1/\rho ^{2}$ moving in a potential $\tilde{V}( \rho
) =\omega ^{2}\rho ^{2}/2-\rho $, the corresponding Schr\"{o}dinger
equation is transformed into an ef\/fectively radial Coulombic problem. Yet,
if the same PDM particle is moving in $\tilde{V}( \rho )
=a^{2}\rho ^{4}/8-d\rho ^{2}/2$, then the Schr\"{o}dinger equation collapses
into a radial harmonic-oscillator problem. In both cases, the exact
solutions can be inferred from the known textbook ones. One observes obvious
dege\-ne\-ra\-cies associated not only with the magnetic quantum number $m$ but
also associated with the ordering ambiguity parameters $\alpha $ and $\gamma
$ (but not $\beta $).

Finally, it should be noted that the applicability of the attendant
methodical proposal can be extended to deal with non-hermitian Hamiltonians
as well (cf., e.g.,~\cite{28} and related references therein).

\vspace{-2mm}

\pdfbookmark[1]{References}{ref}
\LastPageEnding

\end{document}